\documentclass[aps,prd,tightenlines,preprint]{revtex4-1}
\usepackage{graphics,psfrag,amsmath}
\usepackage{epic,eepic}
\usepackage{dcolumn}
\usepackage{bm}
\usepackage{lmodern}
\usepackage{color} 
\usepackage{epsfig}
\usepackage{latexsym}
\usepackage{pstricks}
\usepackage{bbold}
\usepackage{amsmath}
\usepackage{stmaryrd}

\newcommand{\be}{\begin{equation}}
\newcommand{\ee}{\end{equation}}
\newcommand{\bear}{\begin{eqnarray}}
\newcommand{\eear}{\end{eqnarray}}
\begin{document}

\title{Vacuum energy and trace anomaly}
\author{Taekoon Lee}
\email{tlee@kunsan.ac.kr}
\affiliation{Department of 
Physics, Kunsan National University, Kunsan 54150, Korea}


\begin{abstract}
Concerning the trace anomaly in field theory
a nonvanishing vacuum energy breaks the scale symmetry 
as well, in addition to the usual
 beta function dependent term,
requiring a unit operator in the trace anomaly. 
This additional term is also necessary in 
quantum chromodynamics (QCD)  
to cancel the inherent ambiguity in the
gluon condensate. The inseparability of the gluon
condensate effect from the perturbative contribution 
to the vacuum energy renders it impossible
to isolate the gluon condensate effect   on
 the cosmological 
constant. 

\end{abstract}  \pacs{}  

\maketitle

The scale invariance of a quantum field theory is generally broken by 
renormalization, resulting in the trace anomaly, which in a
massless gauge theory is given by \cite{anomaly1,anomaly2}
\bear 
\theta^{\mu}_{\mu}= \frac{\beta(g)}{2g}G^{2}\,,
\label{anomaly}
\eear{}
where $G^{2}=(G_{\alpha\beta})^{2}$, 
$G_{\alpha\beta}$ is the field strength
 tensor,  $\theta_{\mu\nu}$ is
the energy-momentum density tensor,  and $\beta(g)$ is the beta function of the 
gauge coupling $g$.
It shows that the matrix element of the trace of the energy-momentum tensor
between the vacuum and a state of 
two gauge bosons is nonvanishing. Recall that
 the trace anomaly was originally obtained by computing the matrix element.
There is however another matrix element that may not vanish: the vacuum 
expectation value of the trace, 
which in the four-dimensional spacetime is 4 times the vacuum energy
density.   A nonvanishing vacuum energy 
suggests that
$ \theta^{\mu}_{\mu}$ would have a unit operator in addition to
the $G^{2}$ operator. 
In fact Eq. (\ref{anomaly}) is peculiar in that 
the vacuum energy is entirely given 
by the nonperturbative condensate 
of $G^{2}$, with no hint of perturbative contribution.
The additional unit operator can be expected to
 account for the missing perturbative contribution, 
 as $G^{2}$ mixes with the unit operator under
renormalization. 
Therefore, the trace anomaly should be written as
\bear{}
\theta^{\mu}_{\mu}= \frac{\beta(g)}{2g}
G^{2}+4\epsilon_{0}(g)\mathbb{1}\,,
\label{anomaly1}
\eear
{}
where $\epsilon_{0}(g)$ denotes the 
perturbative vacuum energy density.

Unless gravity is involved, ignoring the unit operator 
term is  harmless
as long as
the vacuum energy contribution is subtracted out, that is, only {\it connected}
contribution is taken into account. For example,
the proton mass $M$  in massless QCD can be  given by the trace anomaly \cite{jaffe},
but  it is independent of the unit operator when the vacuum contribution is 
 subtracted out:
\bear{}
2M^{2}&=&\langle P|\theta^{\mu}_{\mu}|P \rangle-
\langle0|\theta^{\mu}_{\mu}|0\rangle \langle P|P\rangle\\  \nonumber{}
&=&\langle P|\frac{\beta(g)}{2g}G^{2}|P \rangle-
\langle0|\frac{\beta(g)}{2g}G^{2}|0\rangle \langle P|P \rangle \\ \nonumber{}
&=&\langle P|\frac{\beta(g)}{2g}G^{2}|P \rangle^{c} \,,
\eear
where $|P\rangle$ denotes the proton state 
and the superscript $c$ denotes the connected 
contribution. 

The addition of the unit operator, however,  resolves a 
 problem that has so far avoided attention, but still 
 important for self-consistency,
 and when the vacuum energy itself is 
 of importance as in gravity.
  The problem with Eq. (\ref{anomaly}) is that the
 condensate of $G^{2}$ in a nonabelian gauge 
 theory is  not a 
 quantity that can be defined unambiguously \cite{david1,david2}, so it
  would imply erroneously  that the vacuum
 energy density is ambiguous. 

Let us consider the anomaly in massless QCD.
 The vacuum energy density $\epsilon$ 
 from Eq.~(\ref{anomaly1}) is  given by
 \bear{}
 \epsilon=\frac{1}{4}\langle0|\theta^{\mu}_{\mu}|0\rangle={}
 \langle0|\frac{\beta(g)}{8g}G^{2}|0\rangle +\epsilon_{0}(g)\,.
\eear
The perturbative series of $\epsilon_{0}$ in  $g^{2}$
is an asymptotic series and its Borel resummation is ambiguous, 
depending on the resummation prescription. 
This ambiguity is to be canceled by that of the
gluon condensate so that the sum would be unambiguous. This
allows one to determine 
the nature of infrared renormalon singularity in the Borel plane
using the renormalization group
 equation of the gluon condensate \cite{mueller}.

To be specific let us consider
 the vacuum energy in lattice QCD without
quarks.
In lattice QCD the vacuum energy can be 
obtained from the average plaquette:

\bear
P(\xi)=\langle 1- \frac{1}{3}\text{Tr} \,\text{U}_{\boxempty}
\rangle=-\frac{1}{N_\boxempty}
\frac{\partial}{\partial\xi}\log Z(\xi)\,,\nonumber
\label{plaquette}
\eear
where $N_\boxempty$ is the  number of plaquettes and
\bear{}
Z=\int \!dU e^{-S(U)} \nonumber
\eear
with
\bear{}
S(U)=\xi\sum_{\boxempty} (1-\frac{1}{3}\text{Tr} 
\,\text{U}_{\boxempty})\,, \nonumber
\eear
and the sum is over all plaquettes,
  $U$  the link variables, and
\bear{}
\xi=\frac{6}{g^{2}}\,. \nonumber
\eear
 For an infinite volume
lattice $Z\sim e^{-\epsilon a^{4} N_{\text {lat}}}$ , where
$\epsilon$ is the vacuum energy density, $a$ 
the lattice spacing, and $N_{\text {lat}}$
is the number of lattice sites, so
we have
\bear{}
P(\xi)=\frac{1}{6}\frac{\partial}{\partial \xi}(\epsilon a^{4})\,,
\eear
where the factor $1/6$ is the ratio 
$N_{\text {lat}}/N_\boxempty$ on a 
four-dimensional lattice.

The average plaquette is given in the form to $O(a^{4})$ \cite{rossi,giacomo}:
\bear{}
P(\xi)=P_{0}(\xi) +\frac{g^{3}}{72\beta(g)}
\langle 0|\frac{\beta(g)}{2g}G^{2}|0\rangle a^{4}\,,
\eear{}
where the beta function is given by
\bear{}
\beta(g(a))=-a\frac{\partial g(a)}{\partial a}\,, \nonumber
\eear
and $P_{0}$ denotes the perturbative contribution:
\bear{}
P_{0}(\xi)=\sum_{1}^{\infty} \frac{c_{n}}{\xi^{n}}\,,{} \nonumber
\eear{}
where the first coefficients are known \cite{rossi}.
Noting that the gluon condensate $\langle0|\beta(g)/(2g)G^{2}|0\rangle$
 is renormalization group (RG) invariant \cite{anomaly2} we get
 the vacuum energy density:
\bear{}
\epsilon&=&\frac{6}{a^{4}} \int P_{0}(\xi) d\xi +
\frac{1}{12a^{4}} \langle0|\frac{\beta(g)}{2g}G^{2}|0\rangle
\int (\frac{g^{3}}{\beta(g)} a^{4}) d\xi \nonumber \\
&=& \epsilon_{0}(g)+
\langle0|\frac{\beta(g)}{8g}G^{2}|0\rangle\,,
\eear
where the perturbative contribution $\epsilon_{0}$
is given by
\bear{}
\epsilon_{0}(g)=\frac{6}{a^{4}} 
\left(c_{1}\log \xi+ \sum_{1}^{\infty}
 \frac{d_{n}}{\xi^{n}}\right) 
\label{series}
\eear
with $d_{n}= c_{n+1}/n$.

As the gluon condensate is RG invariant,
 the condensate is proportional to 
$\Lambda_{\text{QCD}}^{4}
\sim 1/a^{4}e^{-\xi b_{0}}\xi^{\nu}(1+O(1/\xi))$,
where $b_{0}=1/(3\beta_{0})$, $\nu=2\beta_{1}/\beta_{0}^{2}$,
and $\beta_{0},\beta_{1}$ 
are the coefficients of the beta function
\bear{}
\beta(g)=-(\beta_{0}g^{3}+\beta_{1}g^{5}+\cdots){}\,. \nonumber
\eear
An ambiguity of this form in the gluon condensate
 gives rise to the renormalon singularity
\bear{}
\tilde{\epsilon}(b)\sim 1/(1-b/b_{0})^{1+\nu}\,,{}
\label{sing}
\eear
where the Borel transform $ \tilde{\epsilon}(b)$ of the 
perturbative vacuum energy is defined by
\bear{}
\epsilon_{0}(\xi)=\frac{6}{a^{4}}\int_{\cal C}
 e^{-b\xi}\tilde{\epsilon}(b)db\,, \nonumber
\eear{}
and the contour $\cal C$ is along the positive real axis.
The singularity (\ref{sing}) then determines 
the large order behavior of the vacuum energy
coefficients as
\bear{}
d_{n}\sim \Gamma(n+\nu) (3\beta_{0})^{n}\,.\nonumber
\eear
The resummation of the 
asymptotic series for $\epsilon_{0}(\xi)$ then has an 
intrinsic uncertainty of the same magnitude 
as the gluon condensate:
\bear{}
\frac{1}{a^{4}}\Gamma(\bar{n}+\nu){}
\left(\frac{3\beta_{0}}{\xi}\right)^{\bar{n}}\sim 
\Lambda_{\text{QCD}}^{4}\,,
\eear
where $\bar{n}=1-\nu+\xi/(3\beta_{0})$. Thus the
perturbative contribution has a component that is 
comparable to the gluon condensate. Only the sum of the
perturbative contribution and the gluon condensate is 
unambiguous \cite{david2}, which means that
 they cannot be separated from each other.

A consequence of this inseparability is on the 
effect of the gluon condensate in gravity.
The cosmological constant problem---why it is so tiny compared to the
vacuum energy of a field theory---is one of the unsolved, 
most mysterious problems in physics. 
It is clear that 
the effect of the gluon condensate
on the cosmological constant cannot
 be discussed independently of the
perturbative vacuum energy, 
in contrary to   studies \cite{klinghamer1,klinghamer2,brodsky}
that consider
the gluon condensate effect 
 separately from the
perturbative contribution,  or ignoring it entirely; 
Which is not only 
incomplete but also inconsistent.

\begin{acknowledgments}
I am thankful to S. Han for encouragement. 
This research was supported by Basic Science 
Research Program through the National 
Research Foundation of Korea
(NRF), funded by the Ministry of 
Education, Science, and Technology
(2012R1A1A2044543).
\end{acknowledgments}

\bibliographystyle{apsrev4-1}
\bibliography{traceanomaly}

\end{document}